\documentclass[11pt]{article}
\usepackage{latexsym}  % for Symbol fonts
\usepackage{amssymb}
\usepackage{graphicx}
\usepackage{amsmath}
\begin{document}

%%%%%%%%%%%%%%%%%%%%%%%%%%%%%%%%%%%%%%%%%%%%%%%%%%%%%%%

%\preprint{OU-HET-673/2010, MISC-2010-05, TU-871}
\hspace*{11cm} OU-HET-679/2010
%\title{
\begin{center}
{\Large\bf  New U(3) Family Gauge Symmetry and \\
Muonium into Antimuonium Conversion }\footnote{
Invited talk given at The 1st workshop on Intensity Frontier of 
Muon Fundamental Science, KEK, June 10-11, 2010.\ 
To appear in J. of Phys. Conference Series.
} 
  %
  
%}
\vspace{5mm}

{\bf Yoshio Koide}

{\it  Department of Physics, Osaka University,  
Toyonaka, Osaka 560-0043, Japan} \\

E-mail address: koide@het.phys.sci.osaka-u.ac.jp
\end{center}

\begin{abstract}
According to Sumino's idea, 
a family U(3) gauge symmetry is assumed. 
SU(2)$_L$ doublet fields $q_L$ and $\ell_L$ are
assigned to {\bf 3} of U(3), while singlets 
$u_R$, $d_R$, $e_R$ and $\nu_R$ are assigned to
{\bf 3$^*$} of U(3).
Then, current-current interactions with flavor number 
violations of $|\Delta N_{f}|=2$ ($N_{f}$ is 
an individual family number) appear via the family gauge 
boson exchanges.
Since the gauge symmetry model has inevitably been 
brought by Sumino with a specific purpose, 
the gauge coupling constants $g_f$ and the gauge boson mass 
spectrum $m_{fij}\equiv m(A_i^j)$ are not free parameters.  
We estimate $m_{f11} \sim 10^{0-1}$ TeV and 
$m_{f12} \sim 10^{1-2}$ TeV.
As a possible signature of such the flavor number
violating interactions, muonium into antimuonium 
conversion is discussed together with a rare kaon
decay $K^+ \rightarrow \pi^+ + \mu^- +e^+$.
\end{abstract}

%%%%%%%%%%%%%%%%%%%%%%%%%%%%%%%%%%%%%%%%%%%%%%%%%%%%%%%%%%%%
\section{Why do we need a U(3) family gauge symmetry?}

\subsection{Mystery of the charged lepton mass relation}

Investigations of mass spectra have always provided promising 
clues for solving problems in physics: for example, 
the Balmer formula brought the Bohr theory, and 
the Gell-Mann-Okubo mass formula brought the quark model.
Therefore, we may expect that investigation of the 
quark and lepton mass spectra and mixings will also
provide a promising clue to  new fundamental physics.

Meanwhile, in the charged lepton sector, 
we know that an empirical relation~\cite{Koidemass}
\begin{equation}
   K \equiv 
   \frac{m_e +m_\mu + m_\tau}
   {(\sqrt{m_e} + \sqrt{m_\mu} + \sqrt{m_\tau})^2} 
   = \frac{2}{3} 
  \label{K_relation} %(1)
\end{equation} 
is satisfied with the order of $10^{-5}$ 
with the pole masses \cite{PDG08}, i.e. 
\begin{equation}
K^{pole}= \frac{2}{3} \times (0.999989 \pm
0.000014) ,    %(2)
\end{equation}
while it is only valid with 
the order of $10^{-3}$ with the running masses, 
i.e. 
\begin{equation}
K^{run}(\mu)= \frac{2}{3} \times (1.00189 \pm 0.00002) , %(3)
\end{equation}
at $\mu =m_Z$. 
In conventional mass matrix models, ``mass" means not 
``pole mass" but ``running mass." 
Why is the mass formula (\ref{K_relation}) so remarkably satisfied with  
the pole masses? 
This has been a mysterious problem as to the relation (\ref{K_relation}) 
for long years.

\subsection{Sumino mechanism}

Recently, a possible solution of this problem has been
proposed by Sumino \cite{Sumino09PLB,Sumino09JHEP}.
The deviation of $K(\mu)$ from $K^{pole}$ is caused by 
a logarithmic part $m_{ei} \log m_{ei}$ in the radiative correction term
\cite{Arason92}
\begin{equation}
\delta m_{EM} = -\frac{\alpha(\mu)}{\pi} m_{ei} \left( 1+
\frac{3}{4} \log \frac{\mu^2}{m_{ei}^2} \right) .         %(4)
\end{equation}
(Note that the formula (1) is invariant under a  
transformation $m_{ei} \rightarrow m_{ei} (1+\varepsilon_0)$
if the value $\varepsilon_0$ is independent of a family number.)
He considers that a flavor symmetry is gauged, and  
the logarithmic term in the radiative correction to 
the charged lepton masses due to photon is exactly 
canceled by that due to family gauge bosons.
(This does not always mean $m_{ei}(\mu)=m^{pole}_{ei}$.)

%%%%%%%%%%%%%%%%%%%%%%%%%%%%%%%%%%%%%%%%%%%%%%
\begin{figure}[t!]
\begin{center}
\begin{picture}(255,70)
\put(-80,20){\includegraphics[width=155mm]{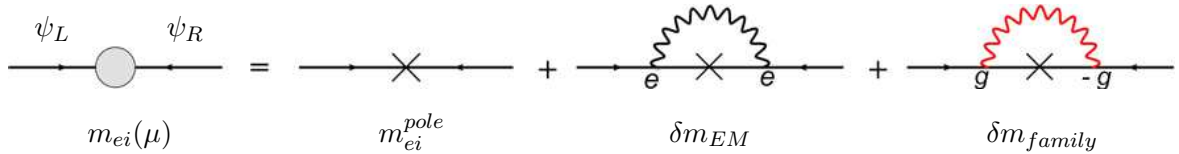}}
\put(-70,40){$\psi_L$}
\put(-20,40){$\psi_R$}
\put(-50,0){$m_{ei}(\mu)$}
\put(60,0){$m_{ei}^{pole}$}
\put(170,0){$\delta m_{EM}$}
\put(290,0){$\delta m_{family}$}
\end{picture}
\end{center}
  \caption{Radiative corrections for charged lepton masses.
The diagram presents $m_{ei}(\mu)=m_{ei}^{pole}$+(radiative
correction due to photon)+(radiative correction due to 
family gauge bosons).}

  \label{sumino_mecha}
\end{figure}
%%%%%%%%%%%%%%%%%%%%%%%%%%%%%

 In order that the Sumino mechanism (cancellation mechanism) works
correctly, a left-handed filed $\psi_L$ and its right-handed
partner $\psi_R$ must be assigned to ${\bf 3}$ and ${\bf 3}^*$ 
of U(3), respectively, i.e. $(\psi_L, \psi_R)=({\bf 3}, {\bf 3^*})$.
Even apart from such a cancellation mechanism, the assignment
$(\psi_L, \psi_R)=({\bf 3}, {\bf 3^*})$ seems to be 
natural also from a point of view of grand unification (GUT) 
scenarios.
If we adopt the conventional assignment in an SU(5) GUT model, 
we must consider
$({\bf 5}^*_L, {\bf 3})+({\bf 10}_L, {\bf 3^*})$ of 
SU(5)$\times$U(3)$_{fam}$, while we can consider 
$({\bf 5}^*_L +{\bf 10}_L, {\bf 3})$ in the new 
assignment $(\psi_L, \psi_R)=({\bf 3}, {\bf 3^*})$. 
Furthermore, we can consider $({\bf 16}_L, {\bf 3})$
in an SO(10) GUT model under the new assignment. 

U(3) family gauge symmetry has, for example, been proposed by
Yanagida \cite{Yanagida79}. 
The assignment $(\psi_L, \psi_R)=({\bf 3}, {\bf 3^*})$ has been
proposed by Appelquist-Bai-Pia \cite{Appelquist06}. 
Their interests are in the possible quark and lepton mass matrix 
forms, and not in searches for visible effects of the gauge bosons. 
A new view in the present U(3) family gauge symmetry model is 
that the gauge coupling constants $g_f$ and the gauge boson mass 
spectrum are not free parameters (see Eqs.(9) and (10) later), 
because the  present gauge symmetry has inevitably been brought by 
Sumino with a specific purpose, as we give a short review 
in the next subsection.
Since the family gauge symmetry proposed by Sumino has the 
energy scale $\Lambda \sim 10^3$ TeV, at which the charged 
lepton mass relation $K=2/3$ is exactly satisfied and the gauge
bosons acquire their masses as shown in Eq.(10), differently 
from other family gauge symmetry models, searches for the 
family gauge boson effects soon become within our reach.

\subsection{Then, what happens?}

The new assignment $(\psi_L, \psi_R)=({\bf 3}, {\bf 3^*})$ can 
induce interesting observable effects.   
In the conventional assignment, a family gauge 
boson $A_j^i$ couples to a current component  
$(J_\mu)_i^j = \bar{\psi}_L^j \gamma_\mu \psi_{Li} +
\bar{\psi}_R^j \gamma_\mu \psi_{Ri}$,
while in the present model, the gauge boson $A_j^i$ 
couples to 
\begin{equation}
   (J_\mu)_i^j 
   = \bar{\psi}_L^j \gamma_\mu \psi_{Li}  
   - \bar{\psi}_{Ri} \gamma_\mu \psi_{R}^j .
   \label{current_general}                  %(5)
\end{equation}
In general, the current-current interactions by
this type of currents cause interactions which violate 
the individual family number $N_f$ by $|\Delta N_f|=2$.
The influence of the flavor number violation is determined 
by the family gauge coupling constant $g_f$ and each 
family gauge boson mass $m_{fij} \equiv m(A_i^j)$. 
Here, for simplicity, the flavor current structure has been 
presented only by a field $\psi$ as a representative of 
the realistic fields 
(quarks $u_i$ and $d_i$ and leptons $e_i$ and $\nu_i$).
For example, the current component $(J_\rho)_1^2$ is
given by
\begin{equation}
   (J_\rho)_1^2 
   = \bar{\mu}_L \gamma_\rho e_{L} 
   - \bar{e}_{R} \gamma_\rho \mu_{R} .
  \label{current_mue}                  %(6)
\end{equation}
This causes a $\mu$ lepton number violation process
$e^-+e^- \rightarrow \mu^- +\mu^-$ through the effective 
current-current interaction
\begin{equation}
   {\cal L}^{eff} 
   = \frac{G_{f12}}{\sqrt2} \left[ 
   \bar{\mu} \gamma_\rho (1 -\gamma_5) e \right]
   \left[ \bar{\mu} \gamma^\rho (1+\gamma_5) e \right] +h.c., 
   \label{Leff1}                      %(7)
\end{equation}
where 
\begin{equation}
\frac{G_{f12}}{\sqrt2} =\frac{g_f^2}{8 (m_{f12})^2} . %(8)
\end{equation}
($m_{f12}=m(A_2^1)$).
Also we may expect flavor violation processes with 
$|\Delta N_f|=2$ 
($N_f$ is an individual family number) such as 
$e^-+e^- \rightarrow \mu^- +\mu^-$, 
$u +e^- \rightarrow c +\mu^-$, $u +u \rightarrow t + t$, 
and so on.

 In the conventional models with $\Delta N_f|=2$ 
 [e.g. the dilepton model \cite{dilepton}],
$g_f$  and $m_{f12}$ are free parameters.
In contrast to the conventional models, the present model has 
the following constraints on the parameters:   

\noindent (i) The coupling constant $g_f$ is related to 
the electric charge $e$ as
\begin{equation}
   \frac{1}{4} g_f^2 = e^2 = g_2^2 \sin^2 \theta_W ,
   \label{gf}  %(9)
\end{equation}
($g_2$ is the SU(2)$_L$ gauge coupling constant) 
in order to work the Sumino mechanism correctly.

\noindent
(ii) The gauge boson masses $m_{fij}$ are related to 
the charged lepton masses $m_{ei}$ as
\begin{equation}
  m_{fij} \equiv m(A_i^j) \propto \sqrt{m_{ei} + m_{ej}} .
  \label{mf}                 %(10)
\end{equation}

The condition (ii) comes from the following reason: 
In this model, Yukawa coupling constants $Y_e^{eff}$ of 
the charged leptons are effectively given by
\begin{equation}
   (Y_e^{eff})_{ij} 
   = \frac{1}{\Lambda^2} \sum_{a=1}^3 
   \langle (\Phi_e)_{ia} \rangle
   \langle (\Phi_e^T)_{aj} \rangle ,
   \label{Yeff}   %(11)
\end{equation}
where $\Phi_e$ is a scalar with $({\bf 3},{\bf 3})$ of
U(3)$\times$O(3) family symmetries.
(Here,  the family U(3)$\times$O(3) symmetries are originated 
in a U(9) flavor symmetry \cite{Sumino09JHEP}, and
only U(3) gauge symmetry can contribute to the radiative
correction of the running masses of charged leptons 
below $\Lambda \sim 10^3$ TeV, at which the charged lepton
mass relation (1) is given exactly.)
In other words, the VEV matrix $\langle \Phi_e \rangle$
is given as 
\begin{equation}
\langle \Phi_e \rangle = {\rm diag}(v_1,v_2,v_3) 
\propto {\rm diag}(\sqrt{m_e}, \sqrt{m_\mu}, \sqrt{m_\tau}).
\end{equation}
[A prototype of such an idea for the charged lepton 
masses is found in Ref.~\cite{K-mass90} related to
the mass formula (\ref{K_relation}).]
Then, the gauge symmetry U(3) is completely broken by
$ \langle \Phi_e \rangle \neq 0$, so that the 
gauge boson masses $m_{fij}$ are related to
the charged lepton masses $m_{ei}$.
In the Sumino model, since the energy scale of the effective 
theory is assumed as $\Lambda \sim 10^3$ TeV, we suppose 
$m_{f13} \sim 10^{2-3}$ TeV and $m_{f12} \sim 10^{1-2}$
TeV.

It is convenient to describe our predictions in terms 
 of a factor $G_{f12}/G_F$, where $G_F$ is the Fermi 
constant $G_F/\sqrt{2}=g_2^2/8 m_W^2$: 
\begin{equation}
   \frac{G_{fij}}{G_F} 
   = 4 \sin^2 \theta_W 
   \left( \frac{m_W}{m_{fij}} \right)^2 
   = \frac{5.98\times 10^{-3}}
   {(m_{fij}\, {\rm [TeV}])^2} .
  \label{G}   %(13)
\end{equation}

%%%%%%%%%%%%%%%%%%%%%%%%%%%%%%%%%%%%%%%%%%%%%%%%%%%%%%%%

\section{Gauge boson mass constraints from rare kaon decays}

\subsection{Current form in the down-quark sector}
\label{startsample}

Next we discuss rare kaon decays. 
Note that, in the present model, the family number $i=(1,2,3)$ 
is defined by $i=(e,\mu,\tau)$ in the charged lepton sector. 
If we assume $i=(1,2,3) \sim (d,s,b)$ for the down-quark sector,
the gauge boson masses $m_{f12}$ can be constrained
by the rare kaon decay searches.  

 In general, the family gauge current forms in the 
quark sectors  depend on quark flavor mixing matrices
 which are given in the  diagonal basis of  the charged 
lepton mass matrix $M_e$.
For simplicity,  we assume that $M_d$ is Hermitian and 
consider only a $d$-$s$ mixing 
\begin{equation}
\left( \begin{array}{c}
d_0 \\
s_0 \\
b_0 
\end{array} \right) = U_d \left( \begin{array}{c}
d \\
s \\
b 
\end{array} \right) = \left( \begin{array}{ccc}
\cos\theta & -\sin\theta & 0 \\
\sin\theta & \cos\theta & 0 \\
0 & 0 & 1
\end{array} \right)  \left( \begin{array}{c}
d \\
s \\
b 
\end{array} \right),   %(14)
\end{equation} 
where 
the down-quark mass matrix $M_d$ is given in the flavor basis 
in which the charged lepton mass matrix $M_e$ is diagonal, and
$M_d$ is diagonalized as $U_d^\dagger M_d U_d = {\rm diag}(m_d, m_s, m_b)$.
In this case, the down-quark current $(J_\mu^{(d)})_1^2$ is given by
\begin{equation}
   (J_\mu^{(d)})_1^2 
   = \bar{s}_L^0 \gamma_\mu d^0_L 
   - \bar{d}^0_R \gamma_\mu s^0_R     \\
   = \frac{1}{2}(\bar{s}\gamma_\mu d - \bar{d}\gamma_\mu s)
   - \frac{1}{2}(\bar{s}\gamma_\mu \gamma_5 d 
   + \bar{d}\gamma_\mu \gamma_5 s) \cos 2\theta     \\
   + \frac{1}{2}(\bar{s}\gamma_\mu \gamma_5 s 
   - \bar{d}\gamma_\mu \gamma_5 d) \sin 2\theta ,
   \label{Jd}             %(15)
\end{equation}
where the first, second and third terms have $CP=-1$, 
$+1$ and $+1$, respectively.
Note that the vector current is independent of the 
mixing angle $\theta$.
(However, this is valid only in the case 
$M_d^\dagger =M_d$.)

As an example of the $s$-$d$ current, let us discuss
a decay of neutral kaon into $e^{\pm}+\mu^{\mp}$.
In Eq.~(\ref{Jd}), only the second term is relevant to
a neutral kaon with
spin-parity $0^-$, which
has $CP=+1$.
Thus, we must identify the second term as $K_S$ (not $K_L$)
in the limit of $CP$ conservation.
Hence, a stringent lower limit of
$m_{f12}$ cannot be extracted from the present 
experimental limit \cite{PDG08} 
$Br(K_L \rightarrow e^{\pm} \mu^{\mp})<4.7 \times 10^{-12}$.

\subsection{Constraints on the gauge boson mass $m_{f12}$}

Instead, the lower limit of $m_{f12}$ can be 
obtained from the rare kaon decays $K^+ \rightarrow \pi^+ 
+ e^{\pm} +\mu^{\mp}$.
The $K \rightarrow\pi$ decay is described by the first term (vector 
currents) in Eq.~(\ref{Jd}), which can be replaced by 
$i(\pi^- \stackrel{\leftrightarrow}{\partial}_\rho K^+)$.
$$ %\begin{equation}
  {\cal L}^{eff} = \frac{g_f^2}{2m_f^2} ( \bar{s}_L \gamma_\rho d_L 
-\bar{d}_R \gamma_\rho s_R + \bar{\mu}_L \gamma_\rho e_L
-\bar{e}_R \gamma_\rho \mu_R) ( \bar{d}_L \gamma^\rho s_L 
-\bar{s}_R \gamma^\rho d_R + \bar{e}_L \gamma^\rho \mu_L
-\bar{\mu}_R \gamma^\rho e_R)  %\nonumber
$$ %\end{equation}
\begin{equation}
 \Rightarrow 
    2 \frac{G_{f12}}{\sqrt{2}} (\bar{s} \gamma_\rho d) 
   (\bar{e} \gamma^\rho \mu - \bar{\mu} \gamma^\rho e)    \\
 \Rightarrow 2 \frac{G_{f12}}{\sqrt{2}} i(\pi^-
   \stackrel{\leftrightarrow}{\partial}_\rho K^+)
   (\bar{e} \gamma^\rho  \mu - \bar{\mu} \gamma^\rho e) .
  \label{Leff2}  %(16)
\end{equation}
Since the effective interaction for $K^+ \rightarrow \pi^0 \mu^+
\nu_\mu$ is given by
\begin{equation}
{\cal L}_{weak} =  \frac{g_2^2}{2m_W^2} V_{us} 
(\bar{s}_L \gamma_\rho u_L) (\bar{\mu}_L \gamma^\rho \nu_{\mu L}) , %(17)
\end{equation}
the ratio $Br(K^+ \rightarrow \pi^+ e^\pm \mu^\mp)/
Br(K^+ \rightarrow \pi^0 \mu^+ \nu_\mu)$ is given by
\begin{equation}
   R 
   = \frac{\bigl[ 2 \cdot (G_{f12}/\sqrt{2}) \bigr]^2}
   {2 |V_{us}|^2 (1/\sqrt{2})^2 (G_F/\sqrt{2})^2} 
   = 67.27 \left(\frac{m_W}{m_{f12}}\right)^4 ,
  \label{R}      %(18)
\end{equation}
in the approximation $m(\pi^+)=m(\pi^0)$ and 
$m(e^-)=m(\nu_\mu)=0$.
Here, we have used the relation (9) for coupling constant
$g_f$.
Thus, the ratio (18) is given only as a linear function 
of the ratio $(m_W/m_{f12})^4$.

The present experimental limits \cite{PDG08} 
\begin{equation}
Br(K^+ \rightarrow \pi^+ e^- \mu^+) <1.3 \times 10^{-11}, 
\ \ \  Br(K^+ \rightarrow \pi^+  \mu^- e^+) <5.2 \times 10^{-10},
\end{equation}
together with $Br(K^+ \rightarrow \pi^0 \mu^+ \nu_\mu)=
(3.35\pm 0.04)\times 10^{-2}$ give lower limits of
the gauge boson mass $m_{f12}$ as shown in Table 1.
Note that the mode $K^+ \rightarrow \pi^+ e^+ \mu^-$
has $|\Delta N_f|=2$, which we are interested in,
while the mode $K^+ \rightarrow \pi^+ e^- \mu^+$ has 
$|\Delta N_f|=0$.

%%%%%%%%%%%%%%%%%%%%%%%%%%%%%%%%%%%%%
\begin{table}[h]  %%%%%%%%%%%%%%%%%%%%%%%%%%%%%%%
\caption{Lower bounds of family gauge boson mass $m_{f12}$ 
estimated from the rare decays $K^+ \rightarrow \pi^+ e^\pm 
\mu^\mp$. Here, we have used $Br(K^+\rightarrow \pi^0 
\mu^+ \nu_\mu) =(3.35\pm 0.04)\times 10^{-2}$.}
\begin{center}
\begin{tabular}{ccc}
\hline %\br
Decay mode &  $K^+ \rightarrow \pi^+  \mu^-e^+$ 
  & $K^+ \rightarrow \pi^+ e^- \mu^+$   \\ 
\hline %\mr
Experiments &  ${Br} <5.2 \times 10^{-10}$
  & ${Br}< 1.3 \times 10^{-11}$
  \\
Lower bound &  $m_{f12} > 21$ TeV
  & $m_{f12} > 52$ TeV
  \\
\hline %  \br
 \end{tabular}
 \end{center}
\end{table}
%%%%%%%%%%%%%%%%%%%%%%%%%%%%%%%%
%%%%%%%%%%%%%%%%%%%%%%%%%%%%%%%%%%%%%%

We can estimate lower bounds of other gauge boson masses, 
$m_{f11}$, $m_{f13}$, etc., from the lower bounds 
of $m_{f12}$ using the relation (10).
The results are listed in Table 2. 
In the present model, the energy scale of the effective
theory $\Lambda$ has been required as $\Lambda \sim 10^3$
TeV. On the other hand, as seen in Table 2, the lower bound
of the mass $m_{f33}$ of the heaviest gauge boson $A_3^3$
is 300 TeV from $K^+ \rightarrow \pi^+ e^- \mu^+$. 
Therefore, the lower bound of each gauge boson listed in
Table 2 seems to be almost near to its upper bound. 
In other words, the mass values given in Table 2 suggest 
that experimental observations 
of family gauge boson effects soon become within our reach.

%%%%%%%%%%%%%%%%%%%%%%%%%%%%%%%%%%%%%%
\begin{table}  %%%%%%%%%%%%%%%%%%%%%%%%%%%%%%%
\caption{{Masses of the gauge bosons $A_1^1$, 
$A_2^1$, $A_3^1$ and $A_3^3$, and their lower bounds 
from rare kaon decays. Their relative sizes are also shown.  }} 
\begin{center}
\begin{tabular}{lllll} \hline
  
  & $m_{f11}$  
  & $m_{f12}$ 
  & $m_{f13}$ 
  & $m_{f33}$ 
  \\ \hline
  Relative sizes
  & $\sqrt{2m_e}$ 
  & $\sqrt{m_\mu +m_e}$
  & $\sqrt{m_\tau +m_e}$ 
  & $\sqrt{2m_\tau}$
  \\
  
  & $0.0981127$ 
  & $1.00000$ 
  & $4.09154$ 
  & $5.78448$
  \\ \hline
  $K^+ \rightarrow \pi^+ \mu^- e^+$
  & 2.1 TeV 
  & 21 TeV 
  & 86 TeV 
  & 120 TeV
  \\
  $K^+ \rightarrow \pi^+ e^- \mu^+$
  & 5.1 TeV 
  & 52 TeV 
  & 210 TeV 
  & 300 TeV
  \\
  \hline
\end{tabular}
\end{center}
\end{table}  %%%%%%%%%%%%%%%%%%%%%%%%%%%%%%%%%
%%%%%%%%%%%%%%%%%%%%%%%%%%%%%%%%%%%%%%%

\section{Muonium into antimuonium conversion}

\subsection{Why important to investigate muonium-antimuonium conversion}

 Exactly speaking, the constraints from kaon rare decays may
 depend on the down-quark mixing structure and some another
hadronic effects, although we suppose that such effects are small.
We would like to test for the new gauge boson effects in the pure 
leptonic processes.   So, the investigation of muonium $M(\mu^+ e^-)$
into antimuonium $\bar{M}(\mu^- e^+)$ conversion is indispensable.

%%%%%%%%%%%%%%%%%%%%%%%%%%%%%%%%%%%%%%%%%%%%%%
\begin{figure}[t!]
\begin{center}
\begin{picture}(255,70)
\put(5,10){\includegraphics[width=85mm]{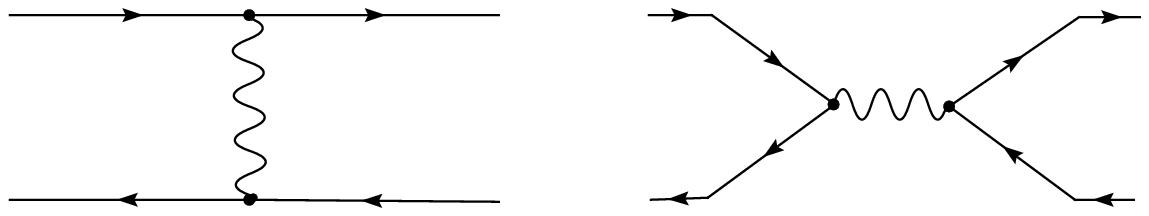}}
\put(5,54){$e^-$}
\put(5,0){$\mu^+$}
\put(64,27){$A_2^1$}
\put(100,54){$\mu^-$}
\put(100,0){$e^+$}
\put(145,54){$e^-$}
\put(145,0){$\mu^+$}
\put(188,40){$A_2^1$}
\put(240,54){$\mu^-$}
\put(240,0){$e^+$}
\end{picture}
\end{center}
  \caption{Diagram of muonium into antimuonium conversion}
  \label{mbarm}
\end{figure}

%%%%%%%%%%%%%%%%%%%%%%%%%%%%%
\subsection{Conversion probability}

The constraint on the gauge boson mass $m_{f12}$ is also  
obtained from a muonium into antimuonium conversion $M(\mu^+ e^-) 
\rightarrow \bar{M}(\mu^- e^+)$.
The total $M\bar{M}$ conversion probability $P_{M\bar{M}}(B)$
under an external magnetic field $B$ 
is generally given by
\begin{equation}
P_{M\bar{M}}(B) = \frac{\delta^2}{{2
[\delta^2+ (E_M-E_{\bar{M}})^2 +\lambda^2]}} ,
\end{equation}
where a mass matrix $M_{mass}$ for the state $(M, \bar{M})$
is given by
\begin{equation}
M_{mass}= \left( 
\begin{array}{cc}
E_{M} & \frac{1}{2} \delta \\
\frac{1}{2} \delta & M_{\bar{M}} 
\end{array} \right) ,               %(20)
\end{equation}
$E_M$ and $E_{\bar{M}}$ are energies of 
muonium $M$ and antimuonium $\bar{M}$, 
$\lambda$ is the muon decay width, and $\delta$ is 
defined by $\langle \bar{M} |H_{M\bar{M}}|M\rangle$ which is
proportional to $(G_{f12}/\sqrt2)/\pi a^3$ ($a$ is the Bohr
radius).  
The case with the effective interaction (7) of a type 
$(V-A)(V+A)$ is effectively
equivalent to a case in the dileton model \cite{dilepton},
and the formulation has been investigated by Horikawa and Sasaki 
\cite{Horikawa96} in details: 
A case with $(V-A)(V+A)$ interaction predicts 
\begin{equation}
P_{M\bar{M}}(0) \simeq \frac{3}{2} \frac{\delta^2}{\lambda^2} ,
\end{equation}
 and 
\begin{equation} 
\delta = -8 \frac{G_{f12}}{\sqrt2} \frac{1}{\pi a^3},
\end{equation} 
so that we obtain
\begin{equation}
P_{M\bar{M}}(0) =1.96 \times 10^{-5} \times \left( \frac{G_{f12}}{G_F}
\right)^2 = 7.01\times 10^{-10}
\frac{1}{m_{f12}}\, {\rm [TeV]}^4 .
\end{equation}
In Table 3, we list predicted values of the predict 
muonium into antimuonium conversion probability 
$P_{M\bar{M}}(0)$ for specific values of 
the family gauge boson mass $m_{f12}$ which correspond 
to the lower limits from
rare kaon decays $K^+ \rightarrow \pi^+ e^\mp \mu^\pm$.
Also, we illustrate behavior of $P_{M\bar{M}}(0)$
versus $m_{f12}$ in Fig.3. 

%%%%%%%%%%%%%%%%%%%%%%%%%%%%%%%%%%%
%%%%%%%%%%%%%%%%%%%%%%%%%%%%%%%%%%%%%
\begin{table}[h]  
\caption{Muonium into antimuonium conversion probability $P_{M\bar{M}}(0)$ for specific values of the family gauge boson
mass $m_{f12}$ which correspond to the lower limits from
rare kaon decays $K^+ \rightarrow \pi^+ e^\mp \mu^\pm$.
}
\begin{center}
\begin{tabular}{ccc}
\hline %\br
Input: $m_{f12}$ &  21 TeV  & 52 TeV \\ 
Prediction: $P_{M\bar{M}}(0)$ &  $3.6 \times 10^{-15}$
  & $9.6 \times 10^{-17}$  \\
\hline %  \br
 \end{tabular}
 \end{center}
\end{table}
%%%%%%%%%%%%%%%%%%%%%%%%%%%%%%%%%%%

\subsection{Present experimental limit on the gauge boson mass $m_{f12}$ }

Present experimental limit \cite{Willmann99} of the total conversion
 probability integrated over all decay times $P_{M\bar{M}}$ is 
\begin{equation} 
P_{M\bar{M}}(B) \le 8.3 \times 10^{-11} \ \  {\rm (90\% CL)\ in} 
B=0.1\ {\rm T}.          %(25)
\end{equation}
Since $S_B(0.1 {\rm T})=0.78$ for the case of $(V-A)(V+A)$ 
\cite{Horikawa96}, where $S_B(B)$ is defined by
\begin{equation}
P_{M\bar{M}}(B) = P_{M\bar{M}}(0) S_B(B) ,  %(26)
\end{equation}
this bound leads to 
$P_{M\bar{M}}(0) \le 1.06 \times 10^{-10}$, and to
$G_{f12}/G_F \le 2.3 \times 10^{-3}$,
so that the lower bound of the mass $m_{f12}$ is given by
\begin{equation}
m_{f12} \ge 19.9\, m_W = 1.60 \ {\rm TeV} .
\end{equation}
The constraint (27) is looser than that from the $K^+$ decay 
given in Table 1. 
However, it should be noticed that  the values in given in 
Table 1 are dependent on the down-quark mixing matrices 
$U_{dL}$ and $U_{dR}$ which are unknown at present differently
from the CKM mixing $V=U^\dagger_{uL} U_{dL}$.  
We would like to emphasize that observations in the pure
leptonic processes are also important independently of 
the bounds from the rare kaon decays.
We expect that future experiments improve this bound (27).

\section{Concluding remarks}

\begin{figure}[h]
\begin{minipage}{16pc}
\includegraphics[width=16pc]{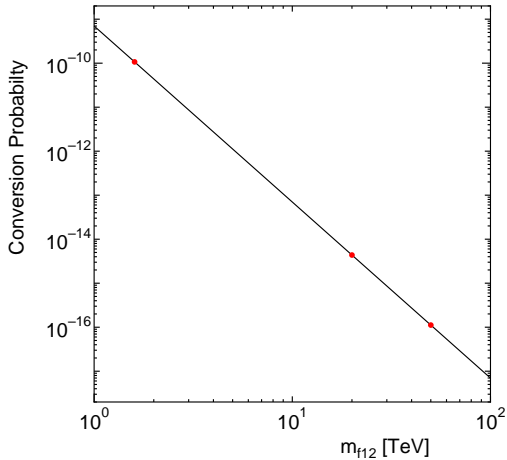}%
\end{minipage}\hspace{2pc}
\begin{minipage}{16pc}\caption{\label{prob}
Conversion probability $P_{M\bar{M}}(0)$ versus 
a gauge boson mass $m_{f12}$.  Points from the left to the right
correspond to the lower limits of the gauge boson mass $m_{f12}$,
1.6 TeV, 21 TeV and 52 TeV,  
which have been estimated from the muonium into antimuonium 
conversion, rare kaon decays $K^+ \rightarrow \pi^+ \mu^- e^+$ 
and  $K^+ \rightarrow \pi^+ e^- \mu^+$, respectively. }
\end{minipage}
\end{figure}

 Although the present lower limit of $m_{f12}$                           
from $M$-$\bar{M}$ conversion experiment is considerably 
looser than those from rare kaon decays as shown in Fig.3.  
However, I again would like to emphasize that, in the present model, 
the family number $i=(1,2,3)$ 
is defined by $i=(e,\mu,\tau)$ in the charged lepton sector. 
Only when we assume $i=(e,\mu,\tau) \sim (d,s,b)$ for the down-quark sector,
the gauge boson masses $m_{f12}$ can be constrained
by the rare kaon decay searches as given in Table 2.  
Therefore, the test for the present family gauge symmetry in the 
$M$-$\bar{M}$ conversion is still indispensable. 
Besides, if we find a positive evidence in the $M$-$\bar{M}$ conversion, 
and if we find that the mass $m_{f12}$ is lower than the lower 
limit of  $m_{f12}$ from the kaon decays, we can obtain an
important clue to the down-quark mixing  $U_d$ (not $V_{CKM}$!) at 
the flavor basis in which the charged lepton mass matrix takes 
a diagonal form.           
We expect that we will soon find a positive evidence in the $M$-$\bar{M}$ 
conversion.

%%%%%%%%%%%%%%%%%%%%%%%%%%%%%%%%%%%%%%%%%%%%%%
%\verb"\ack" 
%\ack %{Acknowledgments}

{\large\bf Acknowledgments}

This work is a part of the work \cite{KSY10} which was done 
in collaboration with Y.~Sumino and M.~Yamanaka. 
The author would like to thank them for their enjoyable 
discussions.

%%%%%%%%%%%%%%%%%%%%%%%%%%%%%%%%%%%%%%%%%%%%%%%%%%%%%%

\end{document}